# Centimeter-Scale Achromatic Hybrid Metalens Design: A New Paradigm Based on Differentiable Ray Tracing in the Visible Spectrum

Qiangbo Zhang, Zeqing Yu, Mengguang Wang, Yiyang Liu, Changwei Zhang, Chang Wang and Zhenrong Zheng[*]

College of Optical Science and Engineering, Zhejiang University, Hangzhou 310027, China

[*] Correspondence: zzr@zju.edu.cn

**Abstract**

Single metalenses are limited by their physical constraints, precluding themselves from achieving high numerical aperture across a wide visible spectral band in large-aperture applications. A hybrid system that integrates a metalens with a refractive lens can address this issue, yet previous designs lacked sufficient flexibility. Here, by reanalyzing the generalized Snell's law, we introduce a new paradigm for the hybrid metalens design based on differentiable ray tracing. Through joint optimization of the phase distribution of the metalens and refractive lens parameters, our system achieves achromatic performance within the broad spectral range of 440-700 nm, with an aperture of 1 cm and an f-number of 1.4. Owing to the differentiable nature of the proposed system, it can be seamlessly integrated as the optical front-end into any differentiable computational imaging system. Our system offers unprecedented opportunities for the advancement of metalenses in innovative optical design and computational imaging domains.

# 1. Introduction

Metalenses, composed of arrays of subwavelength nano-unit cells with lightweight and flat architecture, have emerged as powerful alternatives to traditional refractive lenses due to their ability to freely manipulate the wavefront of light[1–4]. Despite their revolutionary capabilities, metalenses inherently suffer from significant chromatic aberrations. Numerous research efforts have focused on chromatic aberration correction with a single metalens[5–11], invariably subject to a compromise among numerical aperture, bandwidth, and physical dimensions[12,13]. To overcome these limitations, the concept of a hybrid metalens system, which combines metalenses with refractive lenses, has been proposed[14–17]. This innovative system achieves notable chromatic aberration correction over dimensions ranging from millimeters to centimeters. However, within such hybrid systems, the refractive lenses typically remain fixed, lacking joint optimization with the metalenses. This constraint prevents the full potential of the hybrid lens system for chromatic aberration correction.

Recently, refractive lens systems based on differentiable ray-tracing optimization have been introduced[18–22]. By employing a differentiable ray-tracing model, the computational aspects of the refractive optical components become fully differentiable, enabling end-to-end computational imaging when integrated with neural networks for image restoration. Despite such developments in the domain of differentiable ray tracing, integrating metalenses as optical components with refractive lenses within a differentiable ray-tracing framework remains a challenging task.

In this study, we introduce a ray-tracing-based hybrid metalens system, an advanced version of the ***dO*** system[19]—a refractive lens system based on differentiable ray tracing. Our system incorporates the metalens as a fundamental optical component, applying the generalized Snell's law within the entire framework. The ray-tracing procedure of our system is fully differentiable, allowing for the simultaneous optimization of both the metalens and refractive lens. This innovative approach diverges from conventional methods, which solely optimize the metalens while keeping the refractive lens parameters unchanged. Through joint optimization of the phase distribution of the geometric metalens and the aspheric parameters of the refractive lens, we have achieved an achromatic hybrid metalens system across the visible band of 440–700 nm, featuring a 1 cm aperture and an f-number of 1.4. Additionally, the differentiable nature of our framework enables the hybrid metalens system to serve as a front-end optical element, seamlessly coupling with differentiable image-processing components for a broader range of computational imaging applications. The proposed research reveals its potential for bridging the gap between metalenses and refractive lenses and advancing the development of metalenses in the realms of optical design and computational imaging. Source code will be available at https://github.com/Qiangbo100/DeepHybridMetalens.

## 2. Method

### 2.1 Differentiable hybrid metalens ray tracing system

Figures 1a and 1b illustrate the dispersion of light by a geometric metalens and a traditional refractive lens, respectively. When light is incident upon the geometric metalens, the deflection angle increases with longer wavelengths and decreases for shorter wavelengths. Conversely, traditional refractive lenses exhibit the opposite behavior in terms of dispersion. This distinctive characteristic enables the joint design of metalenses and refractive lenses to achieve broadband chromatic aberration correction, as depicted in Figure 1c.

The application of the generalized Snell's law to metalenses introduces a ray-tracing strategy for the integrated design of metalens and refractive lens systems. The schematic diagram of the generalized Snell's law is shown in Figure 2a, with the generalized Snell's law expressed as follows[23]:

$$n_t \sin\theta_t - n_i \sin\theta_i = \frac{\lambda}{2\pi} \frac{\partial \phi(r, \lambda)}{\partial r}, \tag{1}$$

where $n_i$ and $n_t$ denote the refractive indices of the incident and transmission media, respectively, while $\theta_i$ and $\theta_t$ represent the incident and refraction angles of light. The term $\lambda$ is the wavelength of the incident wave, and $\frac{\partial \phi(r, \lambda)}{\partial r}$ symbolizes the phase gradient across the interface. This relation suggests that a refracted beam can be steered in any desired direction by imposing an appropriate gradient of phase at the interface. Furthermore, the unit vector of the outgoing light $\mathbf{d}_t$ can be expressed as:

$$\begin{aligned}\mathbf{d}_t = &\sqrt{1 - \left\|\left[\frac{n_i}{n_t} \cdot (\mathbf{n} \times \mathbf{d}_i) + \frac{\lambda}{2\pi n_t} \frac{\partial \phi(r, \lambda)}{\partial r} \cdot \mathbf{z}\right] \times \mathbf{n}\right\|^2} \cdot \mathbf{n} \\ &+ \left[\frac{n_i}{n_t} \cdot (\mathbf{n} \times \mathbf{d}_i) + \frac{\lambda}{2\pi n_t} \frac{\partial \phi(r, \lambda)}{\partial r} \cdot \mathbf{z}\right] \times \mathbf{n},\end{aligned} \tag{2}$$

where $\mathbf{d}_i$ represents the unit vector of the incident light, $\mathbf{n}$ is the unit normal vector of the metalens,

and $\mathbf{z} = \mathbf{n} \times \mathbf{k}$, with $\mathbf{k}$ denoting the unit vector in the direction of increasing phase gradient. This equation plays a critical role within the hybrid metalens system, as it reveals the direction of refraction for light passing through the metalens. For detailed derivation, please refer to the supplementary material.

The hybrid metalens system based on ray-tracing principles, as illustrated in Figure 2b, begins as light enters the metalens and its direction is altered in accordance with the generalized Snell's law. The light then propagates, interacts with the surface of the aspheric lens and continues propagating according to Snell's law until reaching the sensor plane.

In the ray-tracing process for hybrid metalens systems, due to the presence of complex surface profiles, the intersection points of rays with surfaces must be determined by solving the implicit function $f((x, y, z), \psi) = 0$, where $(x, y, z)$ denotes the position on the intersecting surface and $\psi$ represents the parameters of the optical element. A ray is characterized by its origin $\mathbf{o}$ and unit direction vector $\mathbf{d}$. Using Newton's method, the task turns into finding a positive scalar $t$ that satisfies[18,19]:

$$f((x, y, z), \psi) = f(\mathbf{o} + t\mathbf{d}, \psi) = 0. \tag{3}$$

We utilize the $\mathbf{dO}$ system methodology[19] for an efficient and memory-conserving computation of the scalar $t$. By leveraging the generalized Snell's law alongside Newton's method, we address the problem calculating deflection angles and intersection coordinates, which is essential for ray-tracing based hybrid metalens systems.

## 2.2 Metalens design

Our study utilizes geometric phase metalens with the unit cell shown in Figure 3a, combining silicon dioxide (SiO$_2$) as the substrate and silicon nitride (Si$_3$N$_4$) for nanofins. Unit cell parameters include $L$ = 295 nm, $W$ = 125 nm, $H$ = 800 nm, and $P$ = 350 nm, with rotation angle $\alpha$. With circular polarizers applied to the metalens, the transformation of left-circularly polarized light ($|L\rangle = [1, i]^T$) is determined by specific formulas[1]:

$$\boldsymbol{E}_{out} = \boldsymbol{J}_{meta} |L\rangle = \text{PCE}_\lambda \exp(2i\alpha)|R\rangle, \tag{4}$$

where $\boldsymbol{J}_{meta}$ is the Jones matrix of the unit cell and PCE$_\lambda$ indicates polarization conversion efficiency across wavelengths, as shown in Figure 3b. Equation 4 demonstrates that the phase shift of the geometric metalens is twice the nanofin rotation angle, and is independent of wavelength. This linear correlation is robustly supported by finite-difference time-domain (FDTD) simulations, which consistently showcase this relationship across various wavelengths, as depicted in Figure 3c.

After establishing the relationship between phase shift and rotation angle, we design the metalens phase distribution $\phi$ at the current polar coordinate $r$ as a polynomial expression:

$$\phi(r) = \sum_{i=0}^{n} a_i \left(\frac{r}{R}\right)^{2i}, \tag{5}$$

where $R$ is the radius of the metalens, $\{a_0, \ldots a_n\}$ are optimizable coefficients with $n$ representing the number of terms in the polynomial. In this study, $n$ is set to 7, enabling the optimization of the phase distribution of the metalens through the adjustment of the polynomial coefficients $a_i$. Given that the phase $\phi$ is a function of $r$ and independent of the wavelength $\lambda$, and considering that light is normally incident, with $n_{AIR}$ denoting the refractive index in air, the unit vector of the outgoing light as presented in Equation 2 can be simplified as:

$$\mathbf{d}_t = \sqrt{1-\left(\frac{\lambda}{2\pi n_{AIR}}\frac{\partial \phi(r)}{\partial r}\right)^2} \cdot \mathbf{n} + \left(\frac{\lambda}{2\pi n_{AIR}}\frac{\partial \phi(r)}{\partial r} \cdot \mathbf{z}\right) \times \mathbf{n}. \tag{6}$$

**2.3 Aspherical lens design**

Although our system is capable of optimizing multiple refractive lenses at the same time, this study focuses on employing a single plano-convex lens in conjunction with a metalens to achieve broadband chromatic aberration correction. To increase the optimizable parameters of the refractive lens, we adopt an aspheric surface for the convex side. The expression for the aspheric surface is given as:

$$h(\rho) = \frac{c\rho}{1+\sqrt{1-\alpha\rho}} + \sum_{i=2}^{n} m_{2i}\rho^i, \tag{7}$$

where $c$ denotes the curvature of the lens, $\alpha = (1 + \kappa) c^2$ with $\kappa$ being the conic coefficient, $\rho = r^2$ represents the radial distance square, and $m_{2i}$ are the higher-order coefficients of the aspheric surface. Intersection points with the aspheric lens are determined by Newton's method. Subsequently, Snell's law is employed to ascertain the refraction angles of the emergent rays. In our hybrid metalens system, the aspheric curvature $c$, conic coefficient $\kappa$, and the polynomial coefficients $m_{2i}$ are all optimizable variables. This feature distinguishes our system from previous hybrid metalens designs, which lacked this level of flexibility in the refractive lens parameters.

## 2.4 Loss function

Once parallel light hits the hybrid metalens system, it sequentially passes through the metalens and the aspheric lens before intersecting with the sensor plane at coordinates (x, y, Z), where Z denotes the position of the sensor plane. For the task of chromatic aberration correction, we define the loss function as the root mean square (RMS) of the radii of intersection points across all wavelengths for all traced rays on the sensor plane. If there are N wavelengths, each with M traced rays, the loss function can be expressed as:

$$Loss = \sqrt{\frac{\sum_{i=1}^{NM}\left(x_i^2 + y_i^2\right)}{NM}}. \tag{8}$$

Our system is implemented within the PyTorch framework, where we utilize the ADAM optimizer for gradient descent during the backward propagation. The phase polynomial parameters $a_i$ of the metalens, the aspheric curvature $c$, the conic coefficient $\kappa$, and the polynomial coefficients $m_{2i}$, as well as the distance $d$ between the metalens and the aspheric surface, are all considered as optimizable variables. This provides a significant degree of freedom for our chromatic aberration correction task.

## 3. Results

Our endeavor aims to achieve broadband chromatic aberration correction within the 440-700 nm wavelength range. Initially, we employed the professional optical design software OpticStudio

Zemax to optimize a single aspheric lens, achieving a preliminary design with reasonable chromatic aberration correction. It is crucial to note that, due to inherent limitations, relying on a single aspheric lens does not suffice for optimal chromatic aberration correction. This optimized aspheric lens served as the starting point for refining our system. We positioned the metalens in front of the aspheric lens with its initial phase polynomial coefficients set to zero. Our system then starts operation by simulating incident light at 14 discrete wavelengths between 440-700 nm. For each wavelength, 500 × 500 rays, uniformly generated in Cartesian coordinates, are projected to enter the metalens parallelly, subsequently refracting through both the metalens and aspheric lens and intersecting with the sensor plane. The RMS of the radii of intersection across all wavelengths was computed to define the loss function, which was backpropagated to optimize the system parameters. The descent of the loss function is presented in Figure 4a, illustrating a steady decline that stabilizes as it nears a plateau after 50 epochs. This pattern indicates that our system reliably converges to a robustly optimized solution for the hybrid metalens. Upon completion of the optimization process utilizing gradient descent, we accomplished the design of a hybrid metalens system with a 1 cm aperture and an f-number of 1.4. Detailed optical parameters of the optimized system are listed in the supplementary materials.

The optimized phase profile of the metalens is shown in Figure 4b. Similar to conventional metalenses, the phase is maximal at the center and decreases towards the edges. In our hybrid system, the refractive lens plays a pivotal role in light convergence, thereby reducing the contribution of metalens in focusing tasks. This arrangement results in a more moderate phase shift across the metalens surface.

OpticStudio Zemax possesses the capability to trace rays through large-aperture metalenses, although it cannot optimize metalenses directly. Nevertheless, we can import the phase profile of the optimized metalens and parameters of the aspheric lens into OpticStudio Zemax for validation of our results. Figure 4c displays the RMS radii for an initial single refractive lens and the hybrid metalens across 14 discrete wavelengths ranging from 440-700 nm. It is evident from our system's optimized design that the RMS radii at these wavelengths have significantly decreased. The majority of wavelengths in the hybrid metalens system exhibit RMS values below 10 μm, achieving broadband chromatic aberration correction. Additionally, we calculated the focal lengths (defined here as the distance between the last surface of the aspheric lens and the sensor) across different wavelengths, as shown in Figure 4d. With a single refractive lens, the focal lengths shift between 7.86 to 8.11 mm across the spectrum, clearly leading to substantial chromatic aberration. Conversely, our hybrid metalens system exhibits a much smaller focal length variation, from 7.96 to 8.01 mm, ensuring that light from multiple wavelengths is concentrated onto the same focal plane and significantly mitigating chromatic aberrations.

Figure 5 displays the geometric point spread function (PSF) of the optimized hybrid metalens, calculated by both Zemax and our differentiable hybrid system. The sampling pixel size of the image plane is set to 3.45 μm commonly used by cameras. Despite a methodological divergence in analyzing the phase gradient of the metalens, where we employed a direct differentiation approach as opposed to Zemax's linear interpolation method, the PSFs calculated by our hybrid metalens system closely match those obtained from Zemax. This consistency is crucial within our fully differentiable optimization framework, as the computed PSFs contribute to further analyses of image blurring levels and provide a reliable optical interface for potential subsequent image restoration modules.

To further demonstrate the chromatic aberration correction capability of our hybrid metalens system, we employed the PSFs ($P_\lambda$) to simulate imaging of the 1951 USAF resolution test chart using the following equation:

$$J_{RGB} = \int (I_\lambda * P_\lambda) \cdot S_{\lambda,rgb} d\lambda, \tag{9}$$

where $I_\lambda$ represents the original hyperspectral image (assuming that the grayscale values of the 1951 USAF resolution test chart are the same across different wavelength channels), $S_{\lambda,rgb}$ is the spectral sensitivity function of the sensor (utilizing a Nikon D700 camera in this instance) and $*$ denotes the convolution operator. The results of the imaging simulation are presented in Figure 6. The result of imaging the original 1951 USAF resolution test chart (Figure 6a) using a single initial aspheric lens is displayed in Figure 6b, demonstrating substantial chromatic aberration and severe blur. In contrast, the image produced using our optimized hybrid metalens system, as displayed in Figure 6c, shows virtually no chromatic aberration. Moreover, due to the significantly smaller RMS radius of our system than that of the single initial aspheric lens, the resulting image is markedly clearer and sharper. From the perspective of quantitative image quality metrics, the imaging result of our system achieves an average peak signal-to-noise ratio (PSNR) of 23.81 and a structural similarity (SSIM) of 0.91, markedly surpassing the performance of the single initial aspherical lens.

## 4. Conclusion

In this paper, we introduce a novel ray-tracing based hybrid metalens system that achieves an achromatic performance across the visible band of 440–700 nm, featuring a 1 cm aperture and an f-number of 1.4. This accomplishment is realized through the joint optimization of the phase distribution of the metalens and the aspheric parameters of the refractive lens. Our proposed ray-tracing based hybrid metalens system establishes a bridge for the joint optimization of metalenses and refractive lenses. Furthermore, owing to its differentiability, our system can be integrated into any fully differentiable computational imaging system framework, collaborating with backend image restoration neural networks to enable a broader range of computational imaging applications. Overall, our work contributes to the further development of metalenses within optical design and novel computational imaging frameworks.

**Acknowledgements**

The work is supported by the National Key Research and Development Program of China (2022YFF0705500), National Natural Science Foundation of China (62305092). We thank Dr. Congli Wang for helpful discussions.

**Conflict of Interest**

The authors declare no conflict of interest.

**Author contributions**

Q.Z. and Z.Z. conceived the idea. Q.Z., Z.Y. and C.Z. performed the numerical simulations. All authors contributed to the interpretation and analysis of results. Z.Z. directed the project. All authors participated in paper preparation.

**Data Availability Statement**

The data that support the findings of this study are available from the corresponding author upon reasonable request.

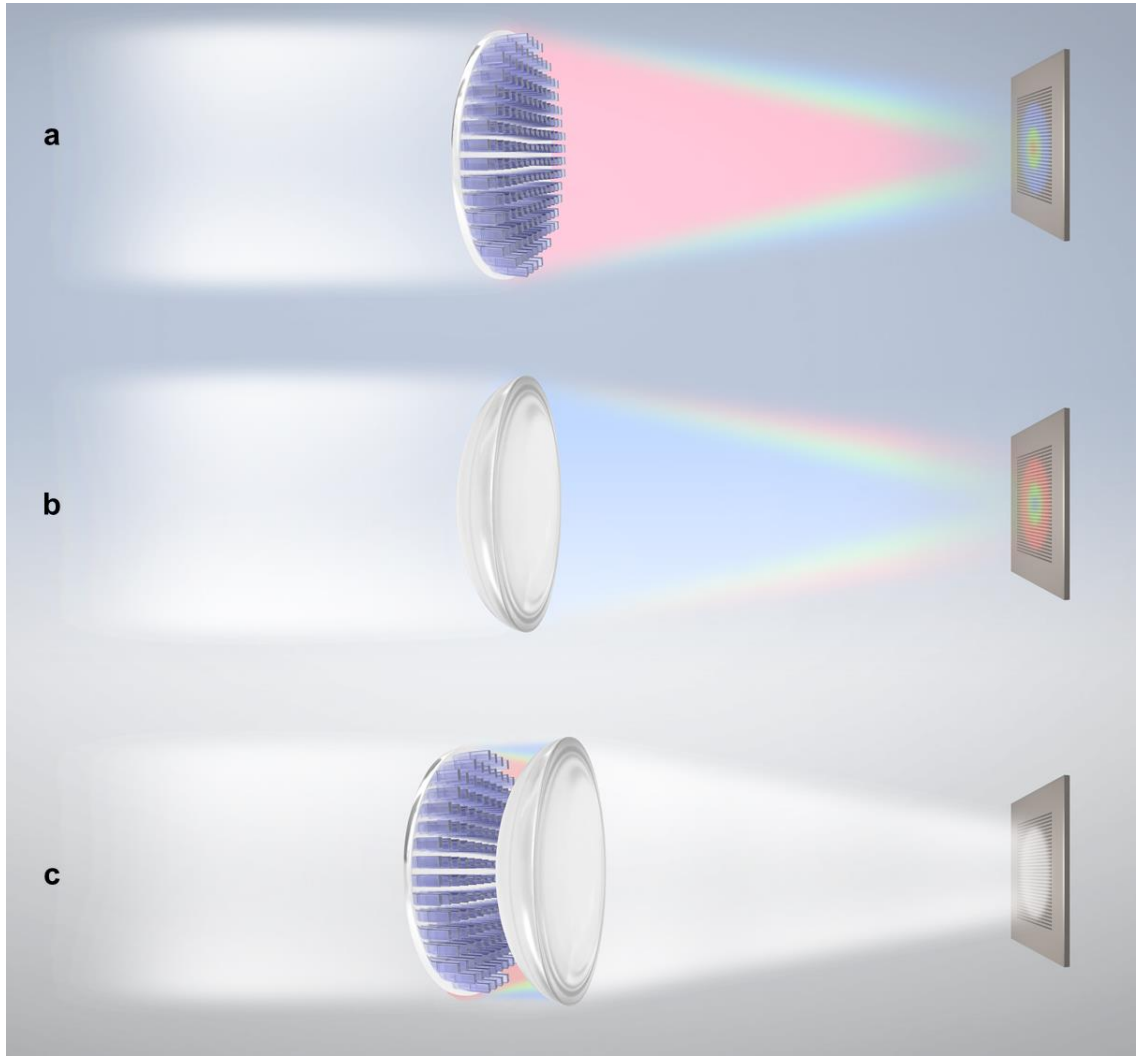

Figure 1. Dispersion of broadband light after passing through (a) the metalens, (b) the refractive lens, and (c) the hybrid metalens system.

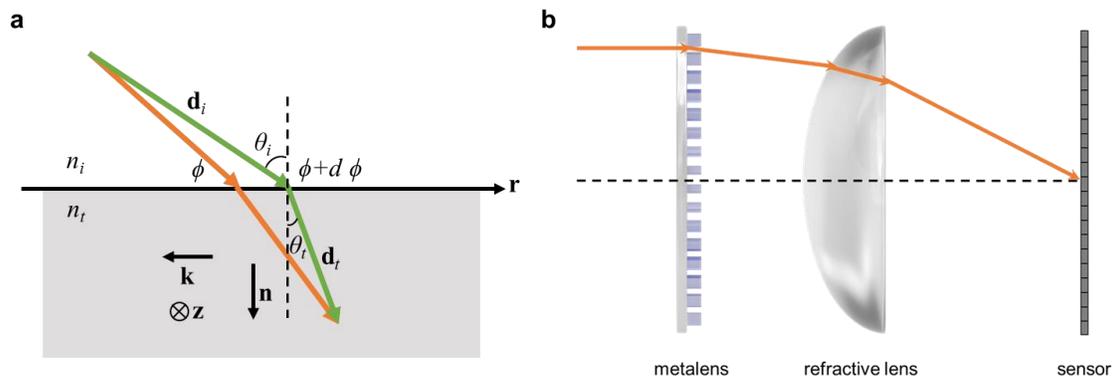

**Figure 2. Schematic diagram of a hybrid metalens system based on ray tracing. a** Schematic diagram of the generalized Snell's law. **b** Schematic diagram of proposed hybrid metalens system.

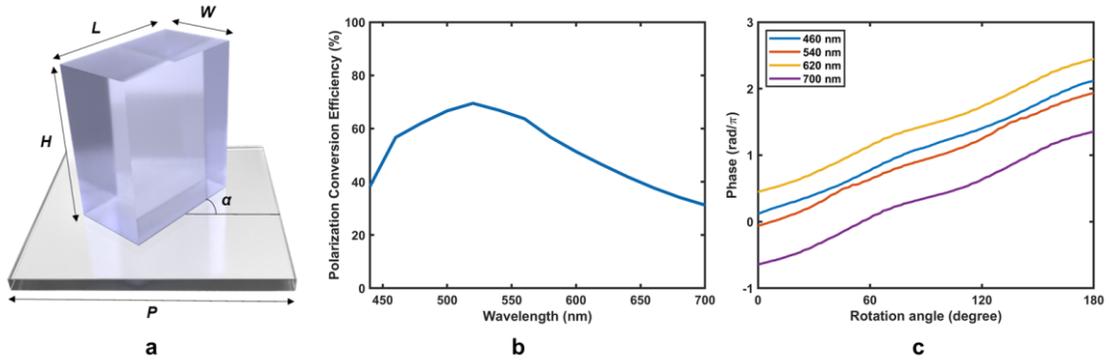

**Figure 3. Characteristics of the metalens unit cell. a** Schematic diagram of the unit cell of the metalens. **b** Polarization conversion efficiency of the metalens unit cell across wavelengths. **c** Relationship between the rotation angle and phase shift of the unit cell at different wavelengths.

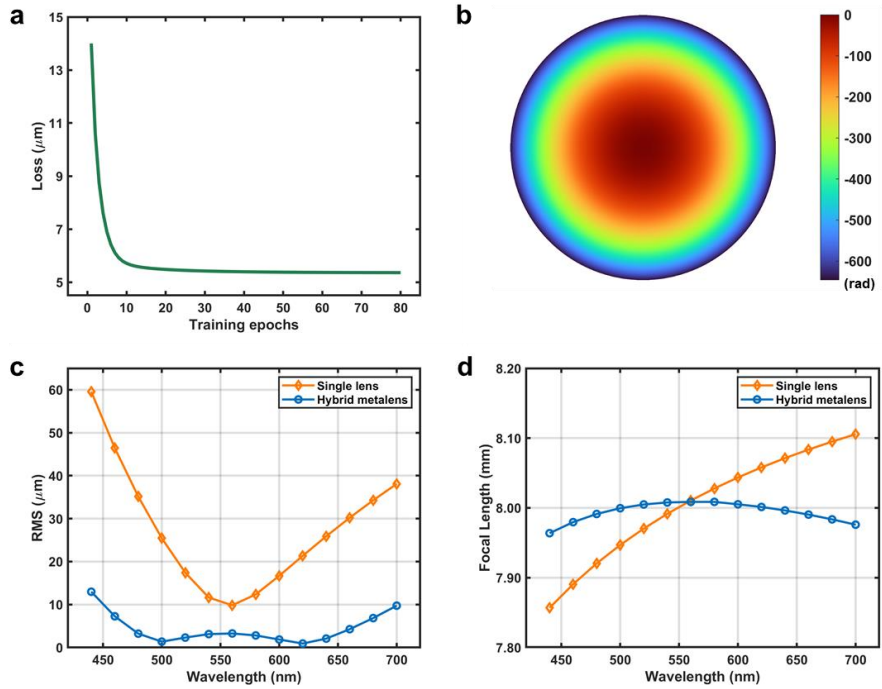

**Figure 4. Optimization results of proposed hybrid metalens system. a** Loss versus number of training epochs. **b** Phase distribution of the metalens after optimization. **c** RMS radius of the single lens and our hybrid metalens system at different wavelengths. **d** Focal length (defined here as the distance between the last surface of the aspheric lens and the sensor) of the single lens and our hybrid metalens system at different wavelengths.

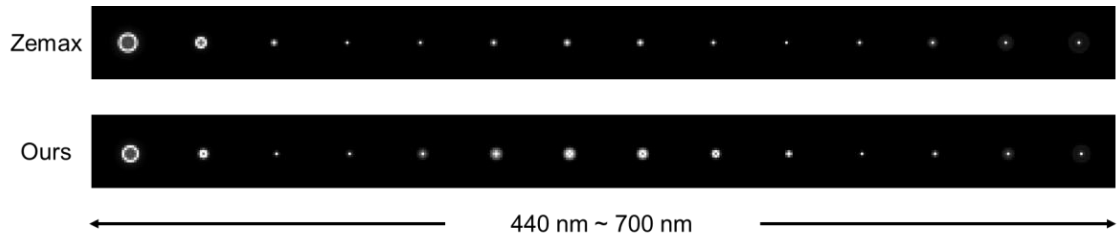

**Figure 5. Comparison of PSFs across 440-700nm wavelengths calculated by Zemax and our hybrid metalens system.**

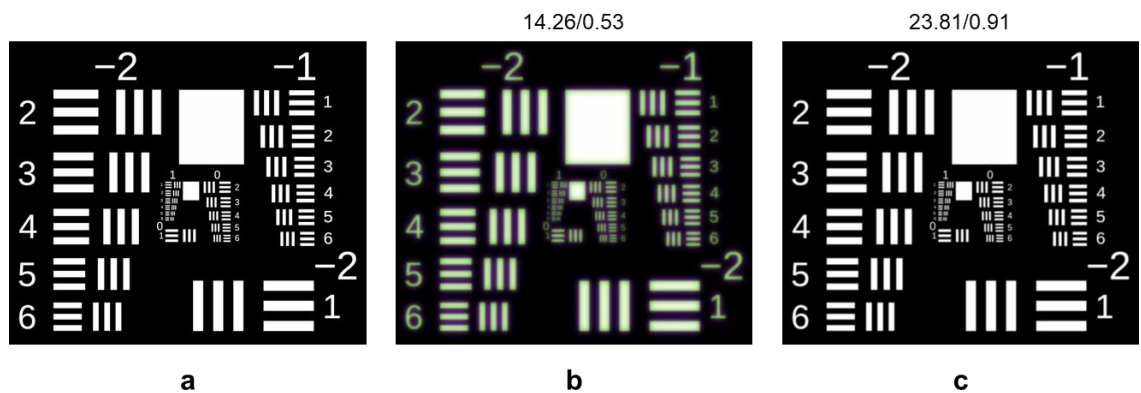

**Figure 6. Imaging characterizations of our hybrid metalens system. a** 1951 USAF resolution test chart (the original picture). **b** simulated images by a single refractive lens. **c** simulated images by the proposed hybrid metalens system. The values above the images indicate the PSNR (dB) and SSIM.